\documentstyle[11pt,aaspp4]{article}

\def\la{\mathrel{\hbox{\rlap{\hbox{\lower4pt\hbox{$\sim$}}}\hbox{$<$}}}}
\def\ga{\mathrel{\hbox{\rlap{\hbox{\lower4pt\hbox{$\sim$}}}\hbox{$>$}}}}

\slugcomment{Submitted to {\it The Astrophysical Journal}}

\begin{document}

\title{Turning to the Dark Side:  Evidence for Circumburst Extinction of Gamma-Ray Bursts with Dark Optical Afterglows}

\author{Daniel E. Reichart\altaffilmark{1,2} and Sarah A. Yost\altaffilmark{1}}

\altaffiltext{1}{Department of Astronomy, California Institute of Technology, Mail Code 105-24, 1201 East California Boulevard, Pasadena, CA 91125}
\altaffiltext{2}{Hubble Fellow}

\begin{abstract}
We present evidence that the majority of rapidly-, well-localized gamma-ray bursts with undetected, or dark, optical afterglows, or `dark bursts' for short, are most likely the result of extinction by dust in the circumburst medium.  First, we show that the dark bursts cannot be explained by a failure to image deeply enough quickly enough:  We fit a variety of brightness distribution models to the optical data, and find that $\approx 57^{+13}_{-11}$\% of all bursts, and $\approx 82^{+22}_{-17}$\% of dark bursts, have afterglows that are fainter than R $= 24$ mag 18 hours after the burst.  Secondly, we show that dark bursts tend to be X-ray and radio faint.  Thirdly, we show that these correlations, and more specifically the optical vs. X-ray and optical vs. radio distributions, can be explained within the framework of the relativistic fireball model if the dark bursts are extinguished by circumburst dust, and the density of the circumburst medium spans many orders of magnitude, from densities that are typical of the Galactic disk to densities that are typical of dense clouds.  Finally, we show that Galactic extinction, host galaxy extinction unrelated to the circumburst medium, and the following high redshift effects -- Lyman limit absorption in the source frame, absorption by the Ly$\alpha$ forest, absorption by excited molecular hydrogen in the circumburst medium, and source-frame extinction by the FUV component of the extinction curve -- might contribute to the number of dark bursts, but only in small numbers.  
\end{abstract}


\section{Introduction}

Optical afterglows have been detected for about 1/3 of the rapidly-, well-localized gamma-ray bursts (e.g., Fynbo et al. 2001; Lazzati, Covino \& Ghisellini 2001).  This data-rich subsample of the rapidly-, well-localized bursts has naturally been the focus of the vast majority of the field's attention and resources over the past four years, but in the end, it is a biased sample.  The nature of the so-called `dark bursts'\footnote{We define `dark burst' to mean a rapidly-, well-localized burst with an undetected optical afterglow.} has only recently become a subject of greater interest, and as might be expected given that these are by definition data-poor events, contradictory initial findings:  Fynbo et al. (2001) find that $\ga 3/4$ of the limiting magnitudes for the dark bursts are brighter than the detected, but faint, optical afterglow of GRB 000630, and conclude that the dark bursts are consistent with a failure to image deeply enough quickly enough.  However, Lazzati, Covino \& Ghisellini (2001), using a smaller sample of dark bursts, find that the distribution of limiting magnitudes for the dark bursts, even if treated as detections, is significantly fainter than the distribution of magnitudes of the detected optical afterglows, and conclude that the dark bursts cannot be explained by a failure to image deeply enough quickly enough.  Lazzati, Covino \& Ghisellini (2001) also find that the X-ray fluxes of the dark bursts are not significantly different than the X-ray fluxes of the detected optical afterglows.  However, we find the opposite below.  

We resolve these issues by applying Bayesian inference, a statistical formalism in which limits can be treated as limits, instead of detections (e.g., Reichart 2001a; Reichart et al. 2001).  We resolve the issue of whether dark bursts can be explained by a failure to image deeply enough quickly enough in \S 2.  Applying the same statistical formalism in \S\S 3 and 4, we show that dark bursts tend to be X-ray and radio faint.  In \S 5, we show that these correlations, and more specifically the optical vs. X-ray and optical vs. radio distributions, can be explained within the framework of the relativistic fireball model if the dark bursts are extinguished by circumburst\footnote{By `circumburst', we mean within the circumburst cloud, which is probably parsecs to tens of parsecs across (Reichart \& Price 2001; see also Galama \& Wijers 2001; Reichart 2001b).} dust, and the density of the circumburst medium spans many orders of magnitude.  We address a number of alternative explanations in \S 6.  We draw conclusions in \S 7.  

\section{Evidence that the Majority of Bursts are Dark}

The samples of Fynbo et al. (2001) and Lazzati, Covino \& Ghisellini (2001) differ in the following ways:  The sample of Fynbo et al. (2001) is nearly complete -- it contains limiting magnitudes for 95\% of the dark bursts preceding GRB 000630 -- whereas the sample of Lazzati, Covino \& Ghisellini (2001) includes limiting magnitudes for BeppoSAX bursts only.  Also, the sample of Lazzati, Covino \& Ghisellini (2001) relies less on GCN Reports, and more on published results.  As BeppoSAX bursts appear to have been deeply imaged more often than bursts detected by other satellites, the different findings of these two papers can be attributed to sample differences, and the practice of comparing limits to detections, which can turn such sample differences into sample biases, as we demonstrate below.

But first, we combine the samples of Fynbo et al. (2001) and Lazzati, Covino \& Ghisellini (2001), keeping only the bursts through GRB 000630 to maintain the completeness of the Fynbo et al. (2001) sample.  As is done in these papers, we scale the data to a common time:  We use a temporal index of --1.4, the median temporal index in Table 1 of Lazzati, Covino \& Ghisellini (2001), and we scale the data to 18 hours after the burst, the median observation time in our combined sample.  In the event of multiple limiting magnitudes for a single burst, we adopt the most constraining.  We plot the combined sample in Figure 1:  The hashed histogram is a binning of the R-band magnitude distribution of the detected optical afterglows.  The unhashed histogram, which is added to the hashed histogram, is a binning of the limiting R-band magnitude distribution of the dark bursts.

Although these distributions appear to be similar, since one is a distribution of limits, they can be consistent only if the vast majority of the dark bursts could have been detected had they only been imaged a little more deeply.  Being this unlucky, consistently, is of course very improbable.  We quantify this improbability with Bayesian inference:  Let $n({\rm R})$ be the normalized R-band magnitude distribution of all of the bursts.  The likelihood function is then given by
\begin{equation}
{\cal L} = \prod_{i=1}^{N}{\cal L}_i,
\label{lf}
\end{equation}
where $N$ is the number of bursts in our combined sample, and  
\begin{equation}
{\cal L}_i = \cases{n({\rm R}_i) & (${\rm R}_i =$ detection) \cr \int_{{\rm R}_i}^{\infty}n({\rm R})d{\rm R} & (${\rm R}_i =$ limit)}
\end{equation}
(e.g., Reichart 2001a; Reichart et al. 2001).  We now consider a wide variety of two-parameter models for $n({\rm R})$:  a Gaussian,
\begin{equation}
n({\rm R}) = \frac{1}{\sqrt{2\pi}b}\exp{\left[-\frac{1}{2}\left(\frac{{\rm R}-a}{b}\right)^2\right]},
\end{equation}
a boxcar,
\begin{equation} 
n({\rm R}) = \cases{0 & (${\rm R} < a$) \cr \frac{1}{b-a} & ($a < {\rm R} < b$) \cr 
0 & (${\rm R} > b$)},
\end{equation}
an increasing power law,
\begin{equation}
n({\rm R}) = \cases{0 & (${\rm R} < 0$) \cr \frac{a+1}{b^{a+1}}{\rm R}^a & ($0 < {\rm R} < b$) \cr 0 & (${\rm R} > b$)},
\end{equation}
and a decreasing power law,
\begin{equation}
n({\rm R}) = \cases{0 & (${\rm R} < b$) \cr -\frac{a+1}{b^{a+1}}{\rm R}^a & (${\rm R} > b$)}.
\end{equation}
We have fitted these models to the data in Figure 1, we plot the best-fit models in Figure 1 (dotted curves), and we list the best-fit parameter values and 68\% credible intervals in Table 1.  Also in Table 1, we list the fractions $f_{ALL}$ of all bursts, and $f_{DARK}$ of dark bursts, that have afterglows fainter than R $= 24$ mag 18 hours after the burst, the magnitude scaled to this time of the faintest detected optical afterglow in our sample.  The results are fairly independent of the assumed shape of the brightness distribution:  On average, we find that $\approx 57^{+13}_{-11}$\% of all bursts, and $\approx 82^{+22}_{-17}$\% of dark bursts, have afterglows that are fainter than R $= 24$ mag 18 hours after the burst.  Consequently, the dark bursts cannot be explained by a failure to image deeply enough quickly enough:  As a whole, they are fainter than the detected optical afterglows.

\section{Evidence that Dark Bursts Tend to be X-ray Faint}

Lazzati, Covino \& Ghisellini (2001) argue that while the optical afterglows of the dark bursts are significantly fainter than the detected optical afterglows, their X-ray afterglows are not similarly faint.  We adopt the Lazzati, Covino \& Ghisellini (2001) sample of 2 -- 10 keV afterglow fluxes, through GRB 000630 (\S 2), and add to it the 2 -- 10 keV afterglow fluxes of non-BeppoSAX dark bursts from a review of IAU Circulars and GCN Reports:  In particular, we have added X-ray data for the RXTE bursts GRB 970616 (Marshall et al. 1997), GRB 970815 (Murakami et al. 1997a), and GRB 970828 (Murakami et al. 1997b), where the latter two were observed by ASCA.  As we did in \S 2, we scale the data to a common time:  We again use a temporal index of --1.4, and we scale the data to 9 hours after the burst, the median observation time in our expanded sample.  We plot these data vs. the optical data from our combined sample (\S 2) in Figure 2.  We notice that most of the dark bursts have 2 -- 10 keV afterglow fluxes between $\approx 10^{-13}$ and $\approx 10^{-12}$ erg cm$^{-2}$ s$^{-1}$, while most of the bursts with detected optical afterglows have 2 -- 10 keV afterglow fluxes between $\approx 10^{-13}$ and $\approx 10^{-11}$ erg cm$^{-2}$ s$^{-1}$.   

We use Bayesian inference to estimate the probability that the optical and X-ray data are correlated.  Let $n({\rm R},\log{F})$ be the normalized distribution of the data in the R -- $\log{F}$ plane.  The likelihood function is again given by Equation (\ref{lf}), but with ${\cal L}_i$ now given by
\begin{equation}
{\cal L}_i = \cases{n({\rm R}_i,\log{F_i}) & (${\rm R}_i =$ detection, $F_i =$ detection) \cr \int_{{\rm R}_i}^{\infty}n({\rm R},\log{F_i})d{\rm R} & (${\rm R}_i =$ limit, $F_i =$ detection) \cr \int_{-\infty}^{\log{F_i}}n({\rm R}_i,\log{F})d\log{F} & (${\rm R}_i =$ detection, $F_i =$ limit) \cr \int_{{\rm R}_i}^{\infty}\int_{-\infty}^{\log{F_i}}n({\rm R},\log{F})d{\rm R}d\log{F} & (${\rm R}_i =$ limit, $F_i =$ limit)}
\label{lfi}
\end{equation}
(e.g., Reichart 2001a; Reichart et al. 2001).  In this section, we consider only a single, simple model for $n({\rm R},\log{F})$: a two-dimensional Gaussian rotated an angle $\theta$ in the R -- $\log{F}$ plane; i.e., contours of constant $n({\rm R},\log{F})$ are simply ellipses rotated in the R -- $\log{F}$ plane.  We consider a physically motivated model in \S 5.  We plot the best-fit model (solid ellipse) in Figure 2.  Clearly, a positive correlation ($0 < \theta < \pi/2$) is favored.  One measure of how significantly $0 < \theta < \pi/2$ is the probability that $-\pi/2 < \theta < 0$.  We find that negative correlations are ruled out at the $\approx 2.7$ $\sigma$ confidence level.  This becomes more significant when we consider the physically motivated model.

One might argue that GRB 970828, the poster child of dark bursts (Groot et al. 1998; Djorgovski et al. 2001), dominates the fit.  Consequently, we redo the fit excluding GRB 970828, and the correlation remains (dotted ellipse).  Clearly, the numerous dark bursts with faint X-ray afterglows dominate the fit.

\section{Evidence that Dark Bursts Tend to be Radio Faint}

We also address the question of whether the radio afterglows of the dark bursts are similarly faint.  We have reviewed the literature, and present our sample of radio flux densities and limits for 25 bursts in Table 2.  We constructed this sample in the following way:  If available, we included the $\approx 8.46$ GHz datum closest to 10 days after the burst.  We selected 8.46 GHz to reduce contamination by interstellar scintillations (e.g., Goodman 1997; Frail et al. 1997), and because it is a commonly used frequency.  We selected 10 days to reduce contamination by reverse shocks (e.g., Sari \& Piran 1999; Kulkarni et al. 1999), and again to reduce contamination by interstellar scintillations.  Ten days is also good because it typically proceeds $t_{jet}$ (Frail et al. 2001), and consequently the light curve is either constant or slowly fading until the synchrotron frequency $\nu_m <$ 8.46 GHz (e.g., Sari, Piran \& Halpern 1999; Harrison et al. 1999):  Hence, our results are fairly independent of the choice of 10 days.  
However, many of the undetected radio afterglows did not have limits available at this frequency or time.  If frequencies were available only at $\approx 4.86$ GHz, we scaled the limits to 8.46 GHz using a spectral index of $+2$, the expected spectral index if the self-absorption frequency $\nu_a >$ 8.46 GHz (e.g., Sari, Piran \& Narayan 1998):  A spectral index of $+1/3$ is expected if $\nu_a <$ 8.46 GHz (e.g., Sari, Piran \& Narayan 1998), so we have adopted the more conservative scaling index.  If times were available only preceding 10 days, we scaled the latest limit to 10 days using a temporal index of $+1$, the expected temporal index if the circumburst medium is the wind of a massive star and $\nu_a >$ 8.46 GHz (Chevalier \& Li 2000):  A temporal index of zero is expected if the circumburst medium is the wind of a massive star and $\nu_a <$ 8.46 GHz (Chevalier \& Li 2000), and a temporal index of $+1/2$ is expected if the circumburst medium is constant in density, independent of whether $\nu_a >$ 8.46 GHz or $\nu_a <$ 8.46 GHz (e.g., Sari, Piran \& Narayan 1998), so again we have adopted the conservative scaling index.  Finally, all of the limits in Table 2 have been converted to $\approx 2$ $\sigma$ before scaling.

For two bursts, GRB 970828 (Figure 3) and GRB 990123 (Figure 4), we can do better.  Although no observation after a few days after the burst resulted in a definitive detection, nearly all of the measurements out to $\approx 100$ dy after the burst favor positive flux densities, suggesting that the afterglow is present at late times, but at a faint level.  Kulkarni et al. (1999) found this to be the case for GRB 990123 by binning the late-time light curve.  Here we take another approach:  We simultaneously fit the forward shock model of Sari, Piran \& Halpern (1999) and the reverse shock model of Sari \& Piran (1999) to the data, and, allowing for interstellar scintillations, constrain the flux density of the forward shock 10 days after the burst.  In the case of GRB 970828, we find that $F_{8.46\,{\rm GHz}}(\Delta t = 10$ dy$) = 16^{+23}_{-9}$ $\mu$Jy, with a median value of 27 $\mu$Jy, and we detect the forward shock at the $\approx 3.0$ $\sigma$ level confidence level.  In the case of GRB 990123, we find that $F_{8.46\,{\rm GHz}}(\Delta t = 10$ dy$) = 16^{+9}_{-6}$ $\mu$Jy, with a median value of 18 $\mu$Jy, and we detect the forward shock at the $\approx 3.2$ $\sigma$ level confidence level.  We use the median values in the following fits.

We plot our radio sample vs. our combined optical sample (\S 2) in Figure 5.  We notice that most of the optically dark bursts are also radio dark.  However, we leave it to the analysis that we presented in \S 3 to decide whether these data are correlated:  Indeed, we again find that a positive correlation is favored (solid ellipse), and we find that negative correlations are ruled out at the $\approx 2.6$ $\sigma$ confidence level.  Again, this becomes more significant when we consider a physically motivated model in \S 5.  Also, we again show that exclusion of GRB 970828 does not significantly affect the result (dotted ellipse).  Clearly, the numerous dark bursts with undetected radio afterglows dominate the fit.  

\section{Evidence that Dark Bursts are Extinguished by Circumburst Dust and that the Density of the Circumburst Medium Spans Many Orders of Magnitude}

If the dark bursts were the result of extinction by dust unrelated to the circumburst medium, e.g., dust elsewhere in the host galaxy, one would expect very little correlation in Figures 2 and 5, since the optical properties of the afterglow would be dominated by this unrelated dust, and the X-ray and radio properties of the afterglow would be determined by the physical properties of the relativistic fireball and circumburst medium.  However, if the dark bursts were the result of extinction by circumburst dust, correlations might be expected, since the optical, X-ray, and radio properties of the afterglow would all depend on the density of the circumburst medium (e.g., Sari, Piran \& Narayan 1998; Sari \& Esin 2001; Harrison et al. 2001).  Consequently, in this section, we determine whether the observed correlations, and more specifically the observed optical vs. X-ray and optical vs. radio distributions, can be reproduced theoretically, using the afterglow model of Sari \& Esin (2001) to which we add extinction by circumburst dust (e.g., Reichart 2001a), by varying the density of the circumburst medium while fixing the remaining parameters to canonical values.

However, this is complicated by the fact that the optical flash (e.g., Sari \& Piran 1999; Akerlof et al. 1999) is expected to destroy circumburst dust via sublimation, and the burst and afterglow are expected to change the size distribution of the circumburst dust via fragmentation (e.g., Waxman \& Draine 2000; Galama \& Wijers 2001; Fruchter, Krolik \& Rhoads 2001; Reichart 2001b).  This is further complicated by the fact that the distances to which the optical flash sublimates dust, and the burst and afterglow fragment dust, are functions of grain size:  The sublimation distance is a decreasing function of grain size, while the fragmentation distance is a decreasing function of grain size for large grains and an increasing function of grain size for small grains (e.g., Waxman \& Draine 2000; Fruchter, Krolik \& Rhoads 2001; Reichart 2001b).  In Reichart (2001b), we show that for a canonical distribution of graphite and silicate grain sizes, and a simple fragmentation model, the burst, optical flash, and afterglow burn through $\approx 10L_{49}^{1/2}$ pc of optical depth, where $L_{49}$ is the 1 -- 7.5 eV isotropic-equivalent peak luminosity of the optical flash in units of $10^{49}$ erg s$^{-1}$ (the 1 -- 7.5 eV isotropic-equivalent peak luminosity of the optical flash of GRB 990123; Waxman \& Draine 2000).  Furthermore, this distance is fairly independent of the density of the circumburst cloud, and of the isotropic-equivalent energy of the burst and afterglow.

We now consider under what conditions bursts burn completely through their circumburst clouds.  Frail et al. (2001) show that the bursts for which redshifts have been measured draw upon a fairly standard energy reservoir of $\sim 3\times10^{51}$ erg (for an efficiency at which this energy is converted to gamma rays of $\eta \sim 0.2$; e.g., Beloborodov 2000), and that the wide range of isotropic-equivalent energies that have been implied for these bursts, from $\la 3\times10^{52}$ erg to $\ga 3\times10^{54}$ erg, is primarily the result of a wide range of collimation angles, with half angles ranging from $\la 0.05$ rad to $\ga 0.5$ rad.  If this is indeed the case, one expects a wide range of isotropic-equivalent peak luminosities for the optical flashes of these bursts, and consequently a wide range of optical depth burn distances, ranging from parsecs to several tens of parsecs.  Consequently, since most clouds, including most giant molecular clouds, tend to be less than several tens of parsecs across (e.g., Solomon et al. 1987), strongly collimated bursts likely burn completely through their circumburst clouds, while weakly collimated bursts likely often do not.\footnote{If the circumburst cloud is the nuclear region of an ultraluminous infrared galaxy, as has been proposed by Ramirez-Ruiz, Trentham \& Blain (2001) to explain the dark bursts, even strongly collimated bursts would not burn through the hundreds of parsecs of optical depth that are typical of such regions (e.g., Solomon et al. 1997).  
However, we show in Reichart \& Price (2001) that the limited information that is available on the column densities, measured from absorption of the X-ray afterglow, of the dark bursts in not consistent with this idea.}

Consequently, we consider two cases:  a strongly collimated burst, and a weakly collimated burst.  In the case of the strongly collimated burst, we assume that the burst, optical flash, and afterglow burn completely through the optical depth of the circumburst cloud.  In the case of the weakly collimated burst, we assume that a few parsecs of dust remain.  

Specifically, we adopt the following parameter values:  Using the now-common notation of Sari, Piran \& Narayan (1998) and Sari, Piran \& Halpern (1999), in the case of the strongly collimated burst, we adopt $E = 3\times10^{54}$ erg and $\theta_{jet} = 0.05$ rad, and in the case of the weakly collimated burst, we adopt $E = 3\times10^{52}$ erg and $\theta_{jet} = 0.5$ rad.  For the remaining relativistic fireball parameter values, we adopt $\epsilon_e = 0.3$, $\epsilon_B = 0.01$, and $p = 2.4$, based on the best fit of the model of Sari \& Esin (2001) to the afterglow of GRB 000926 (Harrison et al. 2001).  We have chosen this fit of this model to this afterglow, because it is the first such fit to include (and detect) the inverse-Compton component of the afterglow (see below).  We adopt a redshift of $z = 1$, because it is typical of the redshifts that have been measured for the bursts to date.  Finally, we vary the density of the circumburst cloud from $n = 10^{-1}$ cm$^{-3}$, which is typical of the Galactic disk, to $n = 10^5$ cm$^{-3}$, which is typical of dense clouds, and in the case of the weakly collimated burst, we vary the number of parsecs of circumburst dust that remain from 1 to 10.

Using the same code that Harrison et al. (2001) use, we plot the optical vs. X-ray results in Figure 6, and the optical vs. radio results in Figure 7.  Qualitatively, these curves can be explained as follows:  At a circumburst cloud density of $n = 10^{-1}$ cm$^{-3}$, the ordering of the self-absorption frequency $\nu_a$, synchrotron frequency $\nu_m$, and cooling frequency $\nu_c$ is $\nu_a <$ 8.46 GHz $< \nu_m$ 10 days after the burst, $\nu_m < \nu_{\rm R} < \nu_c$ 18 hours after the burst, and $\nu_c <$ 2 -- 10 keV 9 hours after the burst.  As we increase the density of the circumburst cloud to $n = 10 - 10^2$ cm$^{-3}$, the radio afterglow grows brighter, because the brightness of the spectral peak ($\nu_m$ at these densities), and consequently the brightness of the afterglow between $\nu_a$ and $\nu_c$, increases with increasing density (e.g., Sari, Piran \& Narayan 1998; Sari \& Esin 2001).  
The optical afterglow also grows brighter for this reason, but only until $n = 1 - 10$ cm$^{-3}$.  At higher densities, $\nu_c$, which decreases with increasing density (e.g., Sari, Piran \& Narayan 1998; Sari \& Esin 2001), passes through the optical, resulting in a net decrease in brightness.  Since $\nu_c <$ 2 -- 10 keV for all of these densities, the X-ray afterglow decreases in brightness for the same reason, but only until $n \approx 10$ cm$^{-3}$.  At higher densities, the inverse-Compton component, which grows brighter with increasing density (e.g., Sari \& Esin 2001), dominates.  This is counteracted once again at very high densities (the exact value depends on the adopted value of $E$), because $\nu_a$, which increases with increasing density (e.g., Sari, Piran \& Narayan 1998; Sari \& Esin 2001), passes $\nu_m$ and $\nu_c$, causing the brightness of the spectral peak ($\nu_a$ at these densities), and consequently the brightness of the afterglow at all frequencies, to decrease with increasing density.

As we increase the density of the circumburst cloud beyond $n = 10 - 10^2$ cm$^{-3}$, the radio afterglow decreases in brightness, because $\nu_a$ passes through the radio.  Finally, in the case of the weakly collimated afterglow, as we increase the density beyond $n \approx 10^2$ cm$^{-3}$, the optical afterglow decreases in brightness substantially, since extinction scales as R $\sim n$, where all of the above effects scale only logarithmically with $n$.

We now investigate whether such a two-component model is strongly requested by the data.  First, we investigate whether a general, two-component model is strongly requested by the data, and then we check whether the best-fit model is in at least broad, qualitative agreement with the theoretical expectation.  Again, we apply Bayesian inference.  The likelihood function is again given by Equations (\ref{lf}) and (\ref{lfi}), but we now adopt a simple two-component model for $n({\rm R},\log{F})$:  the sum of two two-dimensional Gaussians, each rotated in the R -- $\log{F}$ plane, where one Gaussian has weight $w$, and the other has weight $1-w$.  We plot the best-fit models (solid ellipses) in Figures 8 and 9.  In the case of the optical vs. X-ray distribution, the two-component model is favored over the one-component model of \S 3 at the $\approx 3.7$ $\sigma$ confidence level (despite the additional model parameters), and it is favored over the correlation-free model of \S 3 at the $\approx 4.1$ $\sigma$ confidence level.  In the case of the optical vs. radio distribution, the two-component model is favored over the one-component model of \S 4 at the $\approx 3.1$ $\sigma$ confidence level, and it is favored over the correlation-free model of \S 4 at the $\approx 3.5$ $\sigma$ confidence level.  Consequently, two-component models are strongly requested by these data, and one-component and correlation-free models are strongly ruled out.  Again, we show that exclusion of GRB 970828 does not significantly affect these results (dotted ellipses).

Finally, we check whether the best-fit models of Figures 8 and 9 are in at least broad, qualitative agreement with the theoretical expectations of Figures 6 and 7.  To this end, we have replotted the best-fit models in Figures 6 and 7, and, frankly, the agreement is remarkable.  Figures 6 -- 9 suggest that if the afterglows of these bursts are similar in their other properties to the afterglow of GRB 000926, the densities of their circumburst clouds span orders of magnitude, from densities that are typical of the Galactic disk to densities that are typical of dense clouds.  In particular, Figures 6 -- 9 suggest that if the afterglows of GRB 970828 and GRB 990123 (R $\approx 20.6$ mag, $F_{2-10\,{\rm keV}} \approx 5.9\times10^{-12}$ erg cm$^{-2}$ s$^{-1}$, $F_{8.46\,{\rm GHz}} \approx 18$ $\mu$Jy) are similar in their other properties to the afterglow of GRB 000926, the densities of their circumburst clouds are of the order $n \sim 10^4 - 10^5$ cm$^{-3}$.  In the case of GRB 990123, this is a prediction that modeling of the afterglow data should be able to either confirm or refute.\footnote{Panaitescu \& Kumar (2001) model the afterglow data of GRB 990123, and find that $n \sim 5\times10^{-4}$ cm$^{-3}$.  However, the data might be better accommodated by a cooling break between the optical and X rays, and a sufficiently dense circumburst medium to introduce an inverse-Compton component to accommodate the X-ray data.}

\section{Evidence Against Alternative Explanations}

We now address a number of alternative explanations:  

{\bf Galactic extinction:}  Lazzati, Covino \& Ghisellini (2001) show that the distribution of Galactic column densities for the dark bursts is consistent with the distribution of Galactic column densities for the bursts with detected optical afterglows, ruling this out as an explanation for the vast majority of the dark bursts.  

{\bf Host galaxy extinction unrelated to the circumburst medium:}  As we discuss in \S 5, if this were the case, one would expect very little correlation in Figures 2, 5, 8, and 9.  However, this does not appear to be the case (\S\S 3, 4, and 5), ruling this out as an explanation for perhaps all but a few of the dark bursts.  

{\bf High redshift effects:}  Lyman limit absorption in the source frame, absorption by the Ly$\alpha$ forest, absorption by excited molecular hydrogen in the circumburst medium (Draine 2000), and source-frame extinction by the FUV component of the extinction curve (e.g., Reichart 2001a) could all result in dark bursts if at sufficiently high redshifts.  However, Lamb \& Reichart (2000) show that unless the burst history of the universe differs dramatically from the star-formation history of the universe, the redshift distribution of the bursts should be fairly narrowly peaked around $z \approx 2$, primarily because there is very little volume in the universe at low and high redshifts.  Weinberg et al. (2001) model the detection efficiency functions of BeppoSAX and IPN, and show that these satellites should detect even fewer bursts at high redshifts, pushing the expected typical redshift for bursts detected by these satellites down to the observed value of about one.  
However, based on their variability redshift estimates of 220 BATSE bursts, Fenimore \& Ramirez-Ruiz (2001; see also Schaefer, Deng \& Band 2001) find that the burst history of the universe might differ dramatically from the star-formation history of the universe, with very many more bursts at high redshifts.  However, using similar variability redshift estimates (Reichart et al. 2001) for 907 BATSE bursts, Reichart \& Lamb (2001) find that only $\approx 15$\% of bursts above BATSE's detection threshold have $z > 5$ (and if the luminosity function is evolving, far fewer bursts below BATSE's detection threshold have $z > 5$).  Consequently, the above high redshift effects probably affect $\la 10$\% of the bursts in our BeppoSAX- and IPN-dominated sample.  Furthermore, Ramirez-Ruiz, Trentham \& Blain (2001) find that the variability redshift estimates for all of the dark bursts for which high resolution BATSE light curves are available have $z < 5$.

{\bf A second class of long-duration bursts with afterglows that are described by a very different parameterization of the relativistic fireball model:}  The postulation of a second class of objects to explain a phenomenon that can be explained more simply is always ad hoc.  However, for the sake of argument, we investigate whether the relativistic fireball model can account for a dark burst like GRB 970828 without calling upon extinction, and if so, what properties the relativistic fireball and circumburst medium must have.  To this end, we have fitted the model of Sari \& Esin (2001) to the following data:  R $> 23.8$ mag 0.14 days after the burst (Groot et al. 1998), $F_{2-10\,{\rm keV}} = 4\times10^{-13}$ erg cm$^{-2}$ s$^{-1}$ 1.2 days after the burst (Murakami et al. 1997b), and $F_{8.46\,{\rm GHz}} = 16^{+23}_{-9}$ $\mu$Jy 10 days after the burst (\S 4).  Given that the number of degrees of freedom is negative, one would expect a $\chi^2$ of zero, and a broadly degenerate solution.  Instead, we find a $\chi^2$ of a few, and the following solution:  $E \sim 3\times10^{52}$ erg, $\epsilon_e \sim 0.3$, $\epsilon_B \sim 10^{-7}$, n $\sim 3\times10^5$ cm$^{-3}$, and $\theta_{jet} \approx \pi/2$ (isotropic).  
These parameter values can be explained as follows:  The high value of $n$ increases the brightness of the inverse-Compton component of the afterglow to accommodate the high X-ray to optical and X-ray to radio relative brightnesses of the afterglow.  However, unless compensated for elsewhere, the high value of $n$ would also (1) push the self-absorption frequency significantly above 8.46 GHz, making the radio afterglow undetectable, and (2) push the jet break time significantly before 1.2 days after the burst, making the X-ray afterglow too faint.  The model compensates for (1) with an extremely low value of $\epsilon_B$, and a value of $E$ that is about an order of magnitude greater than the expectation from Frail et al. (2001), given that the fireball is isotropic.  The model compensates for (2) by making the fireball isotropic.  In any case, although the fitted model cannot be ruled out on physical grounds, it does not fit the data as well as might be expected, and is unavoidably ad hoc.  We note that we have also tried the wind model of Chevalier \& Li (2000) with similar results.

{\bf A second class of long-duration bursts with afterglows that do not arise from relativistic fireballs:}  Once again, the postulation of a second class of objects to explain a phenomenon that can be explained more simply is ad hoc.  However, this postulation is particularly ad hoc:  Whereas in the previous postulation, we abandoned only a canonical parameterization of a successful physical model for a very different parameterization of the same model, in this postulation, we are abandoning all parameterizations of a successful physical model for an unknown physical model.  We include this possibility only for the sake of completeness.

\section{Discussion and Conclusions}

We have shown that about 60\% of all bursts, and about 80\% of dark bursts, have afterglows that are fainter than R $= 24$ mag 18 hours after the burst.  Furthermore, we have shown that, with the exception of perhaps a few bursts, the dark bursts are most likely the result of circumburst extinction, and the density of the circumburst medium probably spans many orders of magnitude, from densities possibly as low as densities that are typical of the Galactic disk to densities probably as high as densities that are typical of dense clouds.  

Frail et al. (2001) we show that the distribution of values of $En^{-1/4}\eta^{3/4}$ spans a range of a factor of only $6.5^{+2.6}_{-1.3}$ (90\% width), where $\eta$ is the efficiency at which the total energy $E$ is converted to gamma rays.  Consequently, the density $n$ of the circumburst medium is constrained to a range of a factor of $\la 10^{3.2^{+0.6}_{-0.4}}$.  This is perhaps an order of magnitude or two smaller than the range of densities that we find in this paper, but consistent given the uncertainties.  However, there is a selection effect that brings these two estimates into better agreement, and that is that the distribution of values of $En^{-1/4}\eta^{3/4}$ is necessarily constructed using only bursts for which the redshift has been measured, and consequently dark bursts are selected against:  Dark bursts for which the redshift has been measured from emission-line spectroscopy of the host galaxy via a radio localization (e.g., GRB 970828; Djorgovski et al. 2001) are rare, probably because optically dark bursts tend to also be radio dark (\S 4).  Since dark bursts probably have $n \ga 10^2$ cm$^{-3}$ (\S 5), the true distribution of values of $En^{-1/4}\eta^{3/4}$ is probably somewhat broader.

Furthermore, since the dark bursts are probably more weakly collimated than the bursts with detected optical afterglows (\S 5), the true distribution of opening angles is probably somewhat shallower than $\theta_{jet}^{-4.5^{+0.5}_{-0.6}}$, the distribution that Frail et al. (2001) find.  Because of this, the true burst rate is probably somewhat greater than $520 \pm 85$ times the observed rate, and consequently neutron star coalescence becomes even more strongly disfavored than what Frail et al. (2001) find.

\acknowledgements
Support for this work was provided by NASA through the Hubble Fellowship grant \#HST-SF-01133.01-A from the Space Telescope Science Institute, which is operated by the Association of Universities for Research in Astronomy, Inc., under NASA contract NAS5-26555.  We are also grateful to Dale Frail, Johan Fynbo, Don Lamb, Alin Panaitescu, and Paul Price for useful discussions and information.

\clearpage

\begin{deluxetable}{ccccc}
\tablecolumns{5}
\tablewidth{0pc}
\tablecaption{R-Band Magnitude Distribution Model Fits and Faint Burst Fractions}
\tablehead{\colhead{Function} & \colhead{$a$} & \colhead{$b$} & \colhead{$f_{ALL}$} & \colhead{$f_{DARK}$}}
\startdata
Gaussian & $23.98^{+0.75}_{-0.57}$ & $2.95^{+0.65}_{-0.49}$ & $0.50^{+0.10}_{-0.08}$ & $0.71^{+0.16}_{-0.12}$ \nl
Boxcar& $17.12^{+0.03}_{-0.95}$ & $34.40^{+8.40}_{-4.42}$ & $0.60^{+0.13}_{-0.14}$ & $0.86^{+0.20}_{-0.21}$ \nl
Increasing Power Law & $5.74^{+2.09}_{-1.73}$ & $26.83^{+1.48}_{-0.45}$ & $0.53^{+0.17}_{-0.12}$ & $0.76^{+0.34}_{-0.18}$ \nl
Decreasing Power Law & $-2.23^{+0.42}_{-0.51}$ & $17.14^{+0.02}_{-0.53}$ & $0.66^{+0.10}_{-0.11}$ & $0.94^{+0.17}_{-0.17}$ \nl
Average & -- & -- & $0.57^{+0.13}_{-0.11}$ & $0.82^{+0.22}_{-0.17}$ \nl
\enddata
\end{deluxetable}

\begin{deluxetable}{cccccc}
\tablecolumns{6}
\tablewidth{0pc}
\tablecaption{Radio Data}
\tablehead{\colhead{GRB} & \colhead{$\nu$ [GHz]} & \colhead{$\Delta t$ [dy]} & \colhead{$F_{\nu}$ [$\mu$Jy]} & \colhead{$F_{\nu}^{scaled}$ [$\mu$Jy]} & \colhead{Ref(s)\tablenotemark{a}}}
\startdata
970228 & 8.46 & 11.0 & $<80$ & $<80$ & 1 \nl
970508 & 8.46 & 9.95 & $610\pm51$ & 610 & 2 \nl
970828 & 8.46 & 10 & $16^{+23}_{-9}$ & 27 & 3, 4 \nl
971214 & 8.46 & 1.37 & $<34$ & $<248$ & 5 \nl
980329 & 8.46 & 9.90 & $179\pm41$ & 179 & 6 \nl
980519 & 8.46 & 14.05 & $142\pm29$ & 142 & 7 \nl
980703 & 8.46 & 5.32 & 965 & 965 & 8 \nl
981220 & 8.46 & 6.30 & $<72$ & $<114$ & 9 \nl
981226 & 8.46 & 8.54 & $169\pm28$ & 169 & 10 \nl
990123 & 8.46 & 10 & $16^{+9}_{-6}$ & 18 & 3, 11 \nl
990217 & 8.6 & 1.09 & $<112$ & $<1028$ & 12 \nl
990506 & 8.4 & 15.66 & $<137$ & $<137$ & 13 \nl
990510 & 8.6 & 9.22 & $127\pm31$ & 127 & 14 \nl
990704 & 4.88 & 1.02 & $<65$ & $<1915$ & 15 \nl
991014 & 8.46 & 1.68 & $<50$ & $<298$ & 16 \nl
991208 & 8.46 & 9.62 & $999\pm50$ & 999 & 17 \nl
991216 & 8.46 & 9.73 & $170\pm72$ & 170 & 18 \nl
000210 & 8.46 & 4.60 & $<55$ & $<120$ & 19 \nl
000301A & 8.46 & 2.75 & $<136$ & $<495$ & 20 \nl
000301C & 8.46 & 12.17 & $483\pm26$ & 483 & 21 \nl
000326 & 4.86 & 1.48 & $<41$ & $<839$ & 22, 23 \nl
000418 & 8.46 & 9.48 & $100\pm25$ & 100 & 24 \nl
000528 & 8.46 & 3.74 & $<80$ & $<214$ & 25 \nl
000615 & 8.46 & 4.87 & $<75$ & $<154$ & 26 \nl
000630 & 4.86 & 1.02 & $<100$ & $<2971$ & 27 \nl
\enddata
\tablenotetext{a}{1. Frail et al. 1998b; 2. Frail, Waxman \& Kulkarni 2000; 3. this paper; 4. Djorgovski et al. 2001; 5. Frail \& Kulkarni 1997; 6. Taylor et al. 1998; 7. Frail et al. 2000a; 8. Frail et al. 1998a; 9. Frail \& Kulkarni 1998; 10. Frail 1999; 11. Kulkarni et al. 1999; 12. Wark et al. 1999; 13. Taylor et al. 2000; 14. Harrison et al. 1999; 15. Rol et al. 1999; 16. Taylor, Frail \& Kulkarni 1999; 17. Galama et al. 2000; 18. Frail et al. 2000b; 19. McConnell et al. 2000; 20. Frail \& Taylor 2000; 21. Berger et al. 2000; 22. Frail 2000a; 23. Frail 2000b; 24. Berger et al. 2001; 25. Berger \& Frail 2000a; 26. Frail, Becker \& Berger 2000; 27. Berger \& Frail 2000b.}
\end{deluxetable}

\clearpage

\figcaption[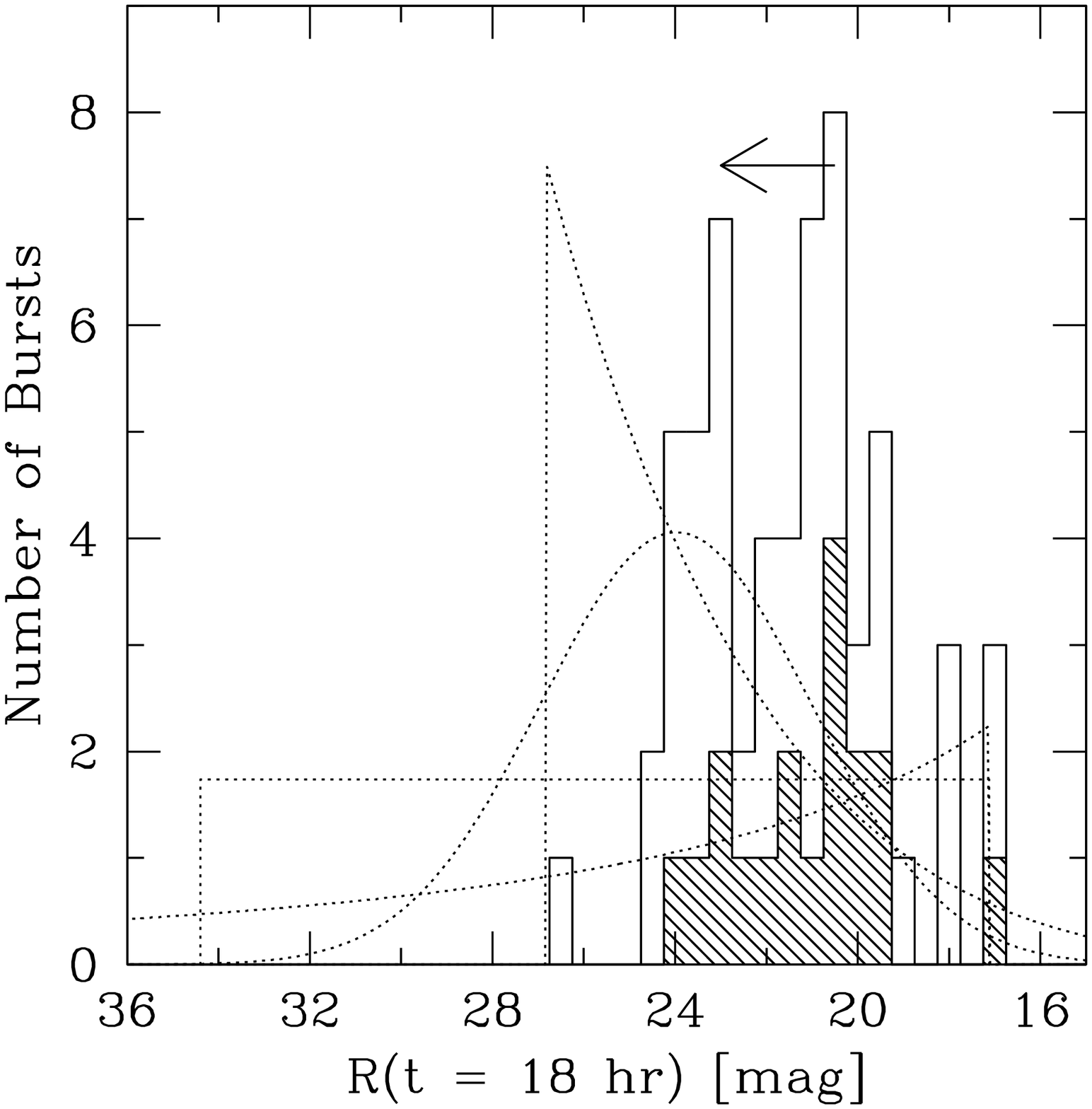]{R-band magnitude distribution of the detected optical afterglows (hashed histogram), and limiting magnitude distribution of the dark bursts (unhashed histogram, added to the hashed histogram), scaled to 18 hours after the burst.  The dotted curves are the best-fit models of \S 2 to these data.  Clearly, most bursts have afterglows fainter than R $\approx 24$ mag 18 hours after the burst, independent of the assumed shape of the brightness distribution.\label{rdist.eps}}

\figcaption[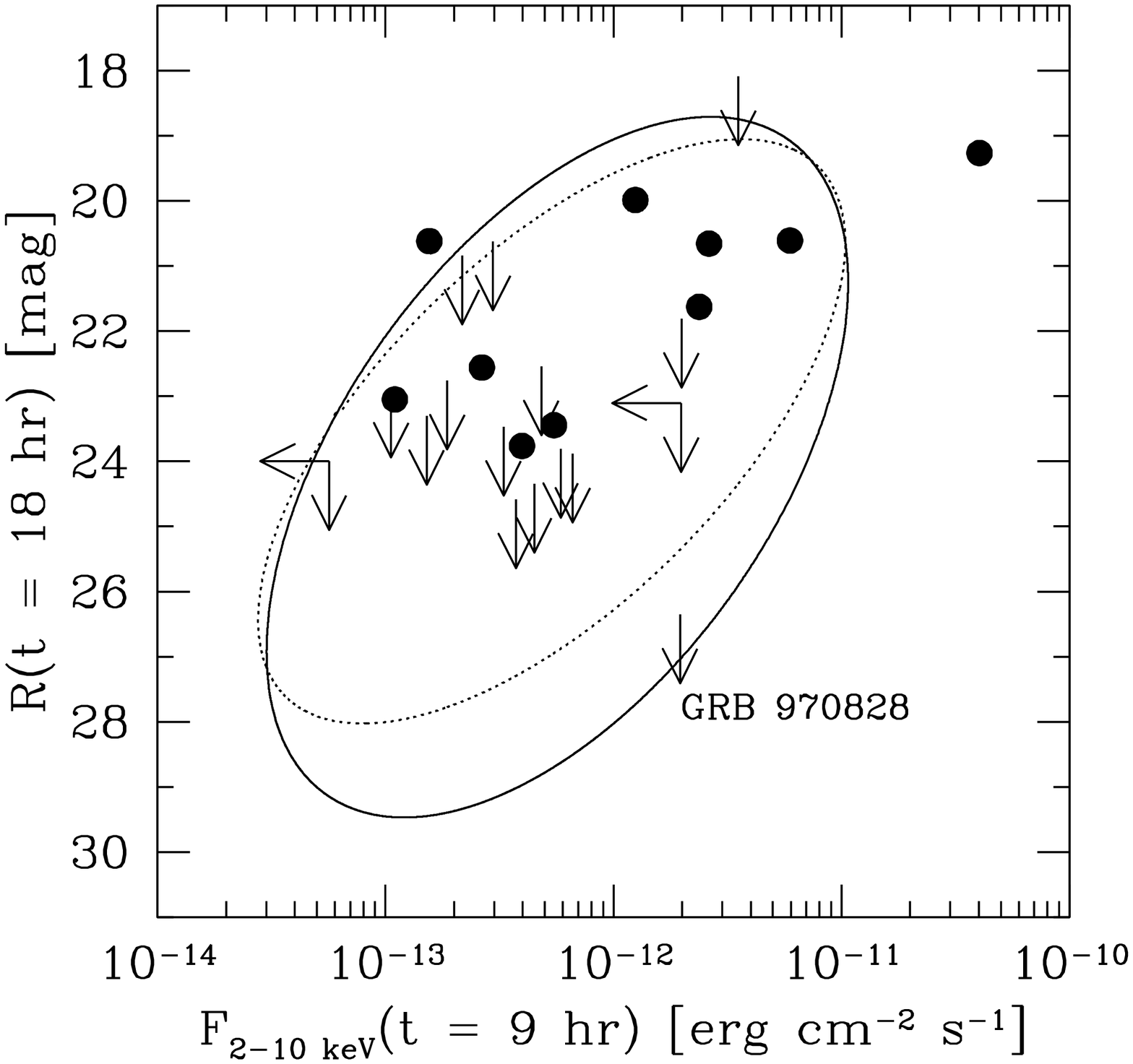]{R-band magnitude or limiting magnitude, scaled to 18 hours after the burst, vs. 2 -- 10 keV flux or limiting flux, scaled to 9 hours after the burst.  The solid ellipse is from the best-fit model of \S 3 to all of the data, and the dotted ellipse is from the best-fit model to all of the data excluding GRB 970828.  The ellipses should encompass $\approx 90$\% of the data.  The optical and X-ray data appear to be correlated, and clearly GRB 970828 does not dominate this correlation.\label{rxdist.eps}}

\figcaption[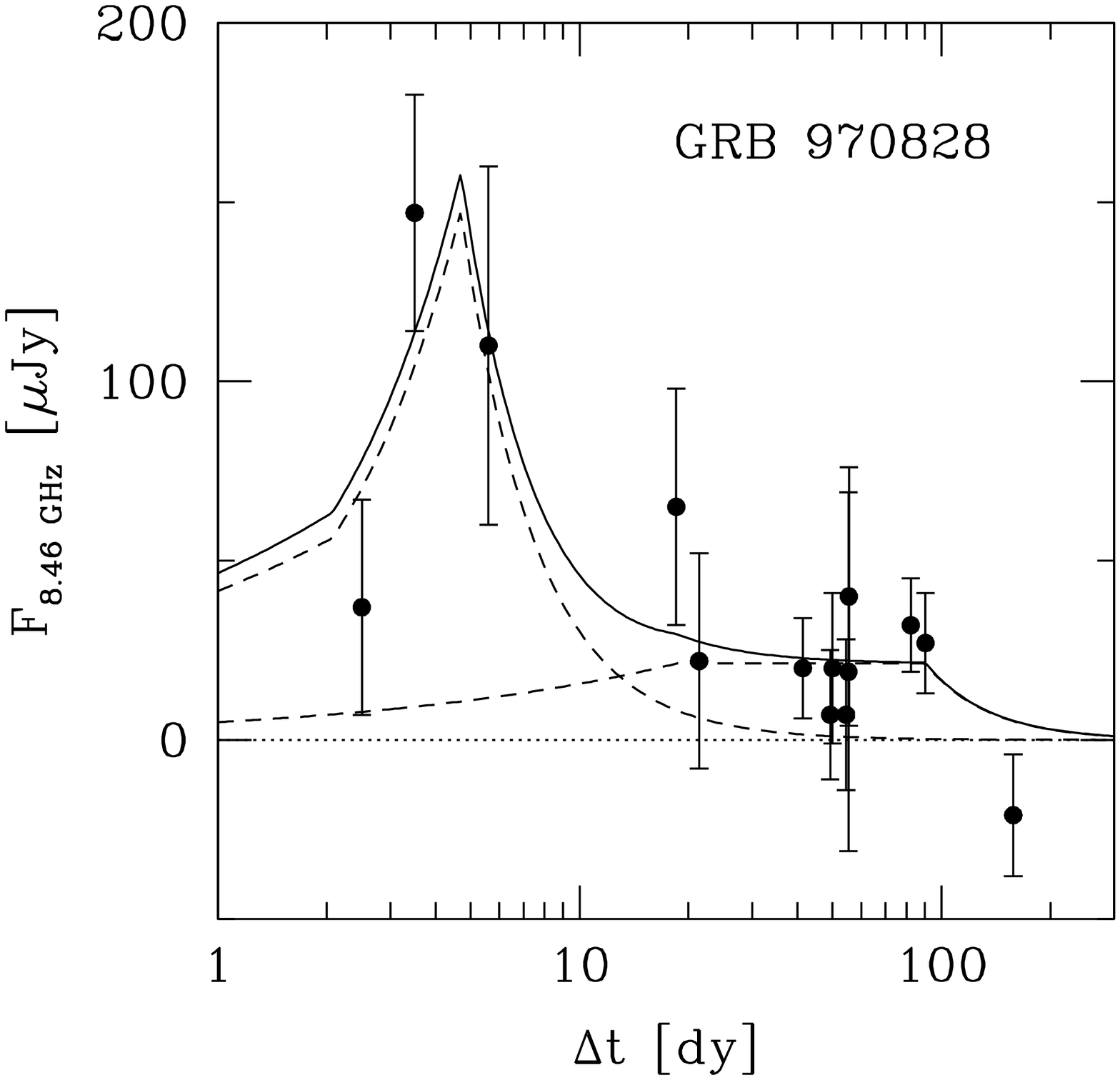]{8.46 GHz light curve of the afterglow of GRB 970828 from Djorgovski et al. (2001).  The dashed curves are the best-fit reverse and forward shock components of the afterglow, and the solid curve is the sum of these components (\S 4).  Although no observation after a few days after the burst resulted in a definitive detection, all of the measurements out to $\approx 100$ dy after the burst favor positive flux densities, suggesting that the afterglow is present at late times, but at a faint level.\label{lc828.eps}}

\figcaption[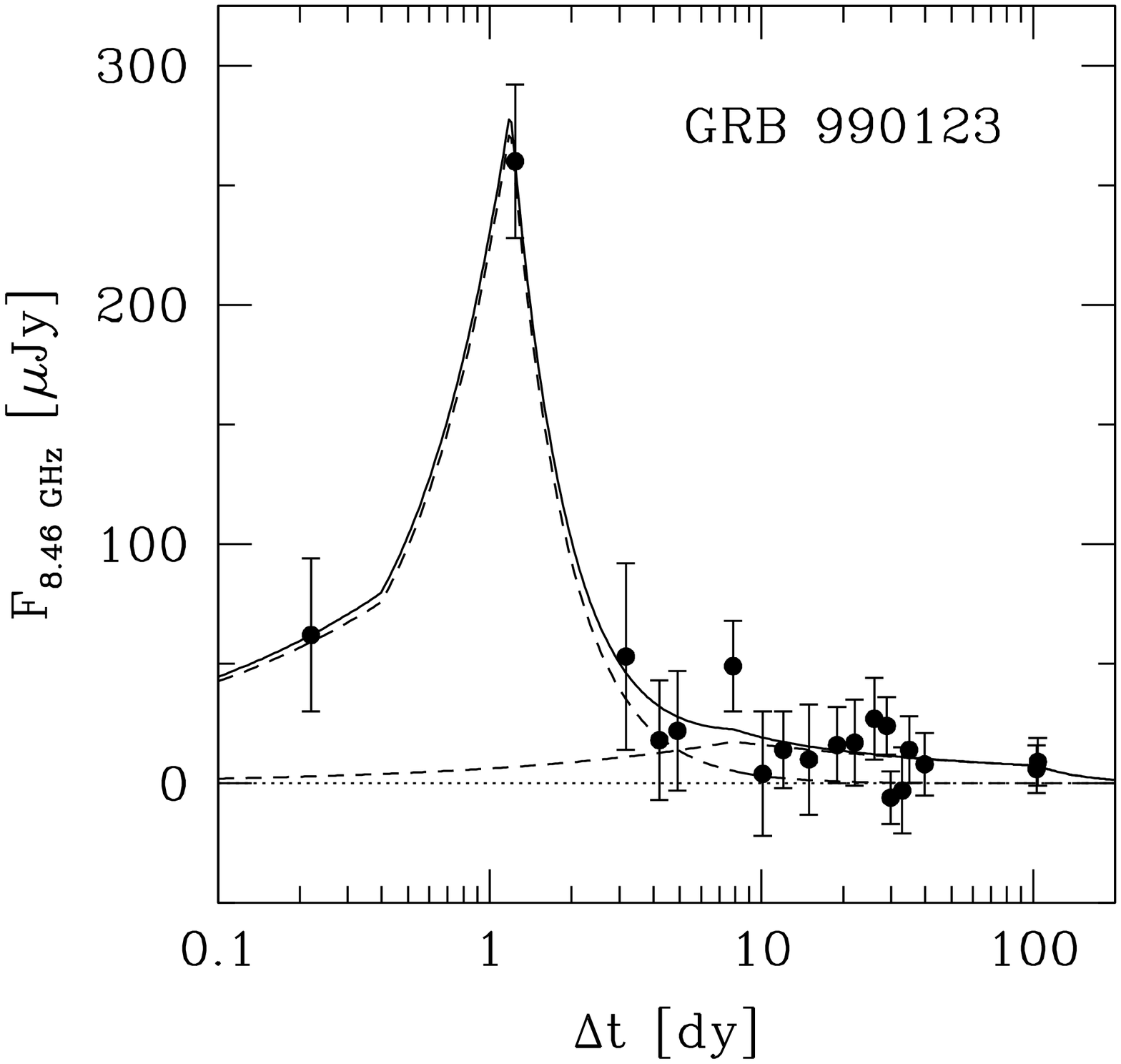]{8.46 GHz light curve of the afterglow of GRB 990123 from Kulkarni et al. (1999).  The dashed curves are the best-fit reverse and forward shock components of the afterglow, and the solid curve is the sum of these components (\S 4).  Although no observation after a few days after the burst resulted in a definitive detection, nearly all of the measurements out to $\ga 100$ dy after the burst favor positive flux densities, suggesting that the afterglow is present at late times, but at a faint level.\label{lc123.eps}}

\figcaption[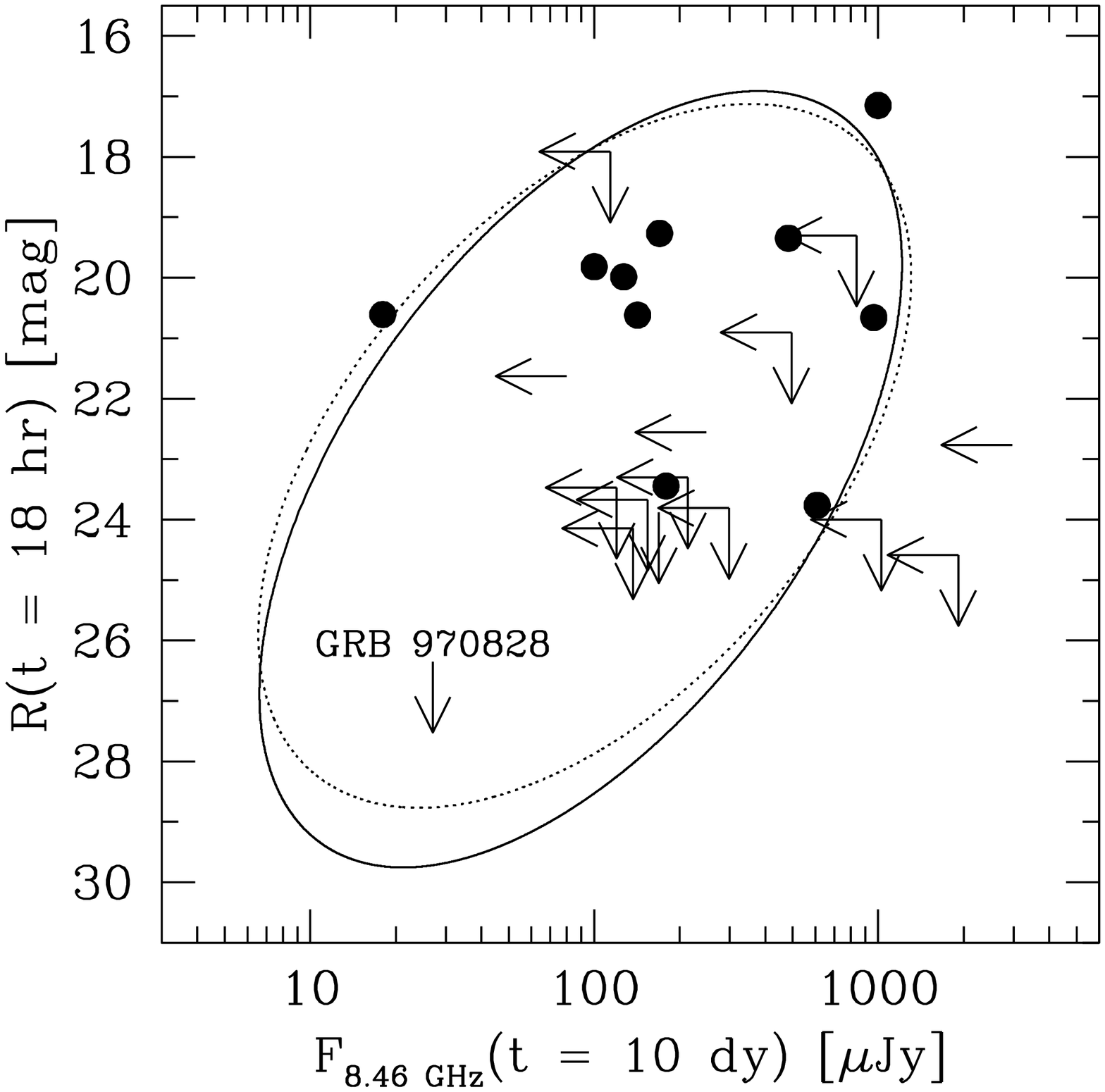]{R-band magnitude or limiting magnitude, scaled to 18 hours after the burst, vs. 8.46 GHz flux density or limiting flux density, scaled to 10 days after the burst.  The solid ellipse is from the best-fit model of \S 4 to all of the data, and the dotted ellipse is from the best-fit model to all of the data excluding GRB 970828.  The ellipses should encompass $\approx 90$\% of the data.  The optical and radio data appear to be correlated, and clearly GRB 970828 does not dominate this correlation.\label{rrdist.eps}}

\figcaption[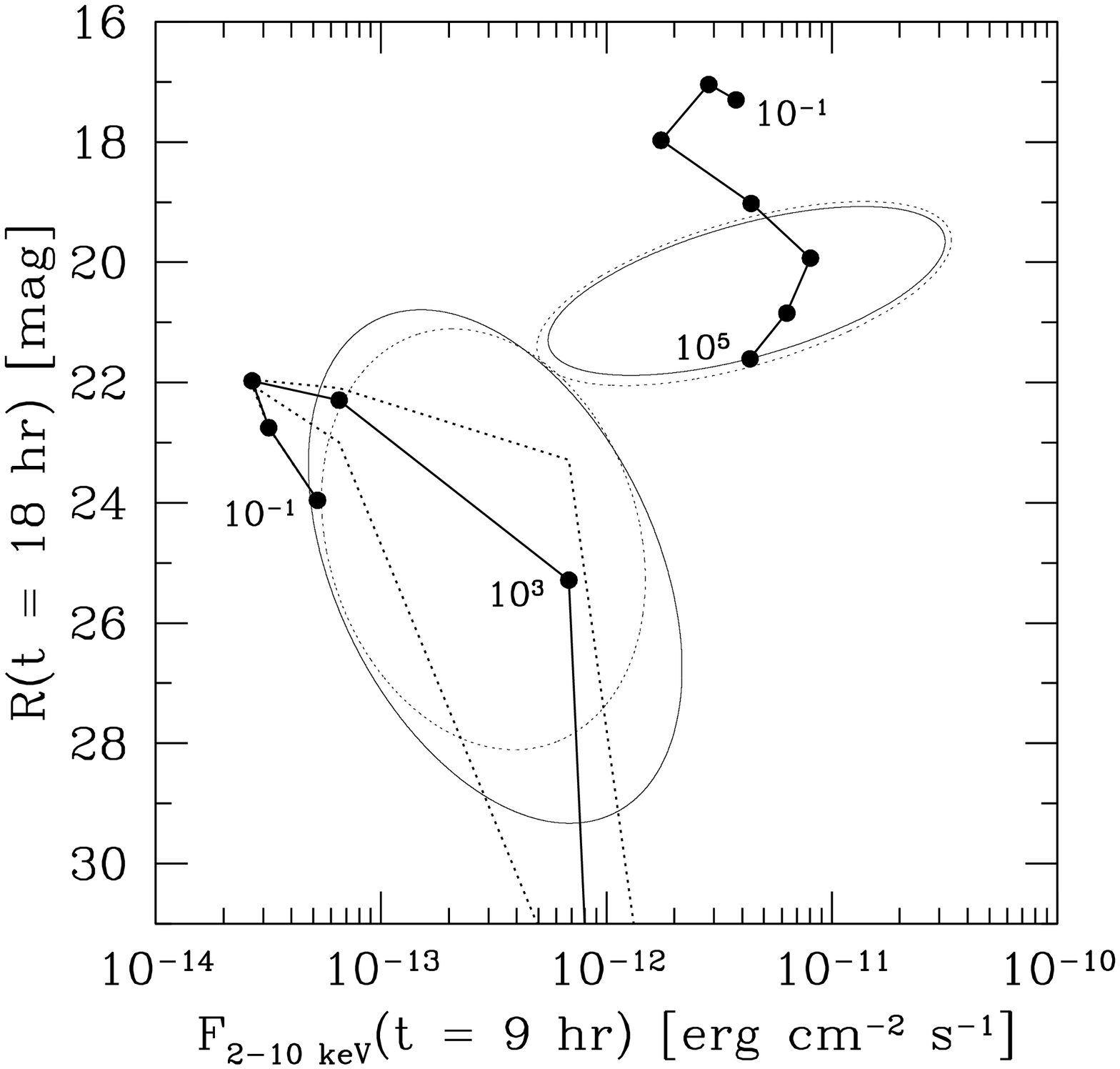]{Theoretical R-band magnitude vs. 2 -- 10 keV flux as a function of circumburst cloud density ($n = 10^{-1} - 10^5$ cm$^{-3}$) for a strongly collimated burst that burns through the dust of its circumburst cloud (solid curve, top right), and for a weakly collimated burst that burns through all but one (top dotted curve, bottom left), three (solid curve, bottom left), and ten (bottom dotted curve, bottom left) parsecs of dust of its circumburst cloud (\S 5).  The ellipses are replotted from Figure 8.\label{rxdist3.eps}}

\figcaption[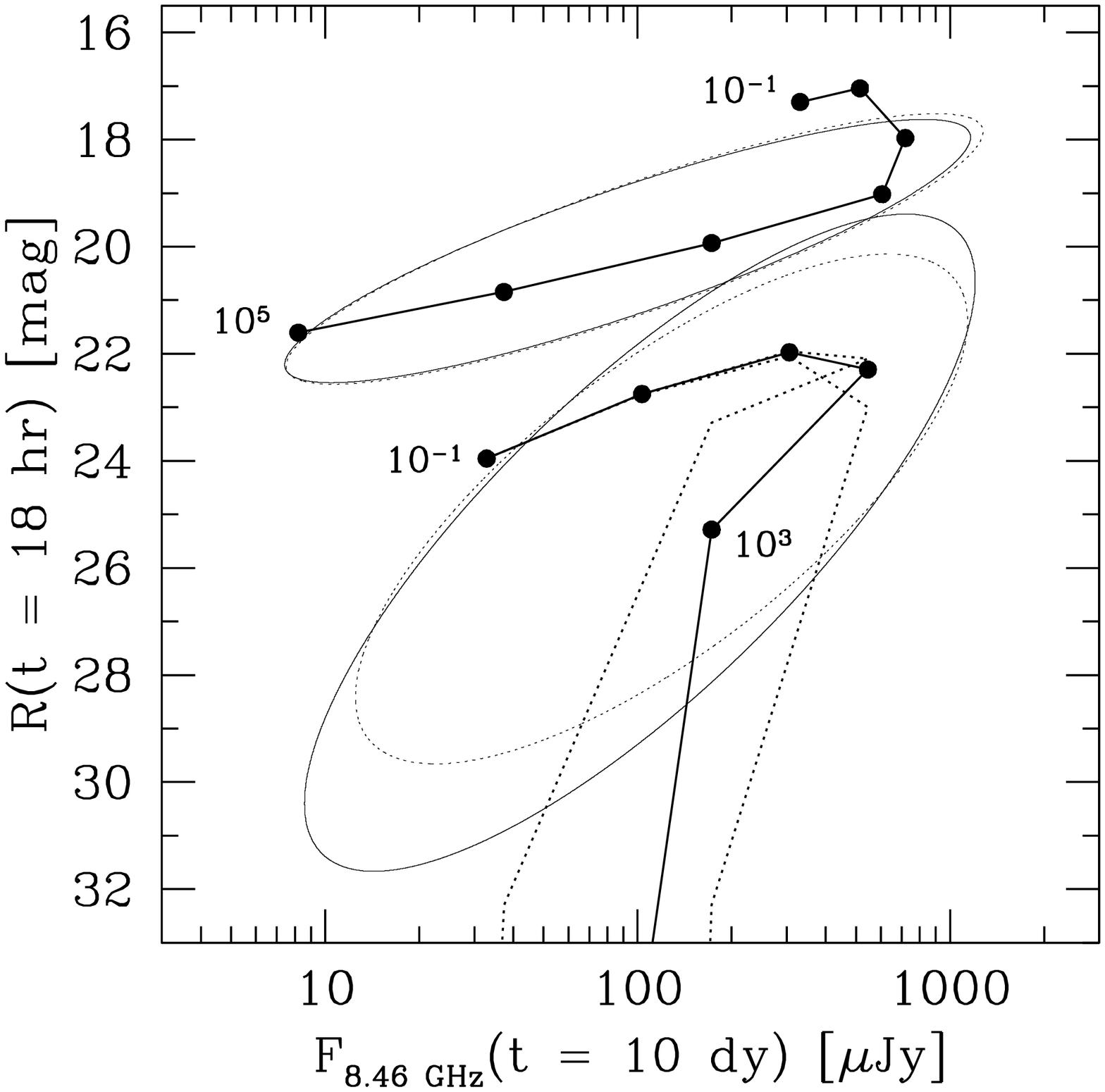]{Theoretical R-band magnitude vs. 8.46 GHz flux density as a function of circumburst cloud density ($n = 10^{-1} - 10^5$ cm$^{-3}$) for a strongly collimated burst that burns through the dust of its circumburst cloud (solid curve, top), and for a weakly collimated burst that burns through all but one (top dotted curve, bottom), three (solid curve, bottom), and ten (bottom dotted curve, bottom) parsecs of dust of its circumburst cloud (\S 5).  The ellipses are replotted from Figure 9.\label{rrdist3.eps}}

\figcaption[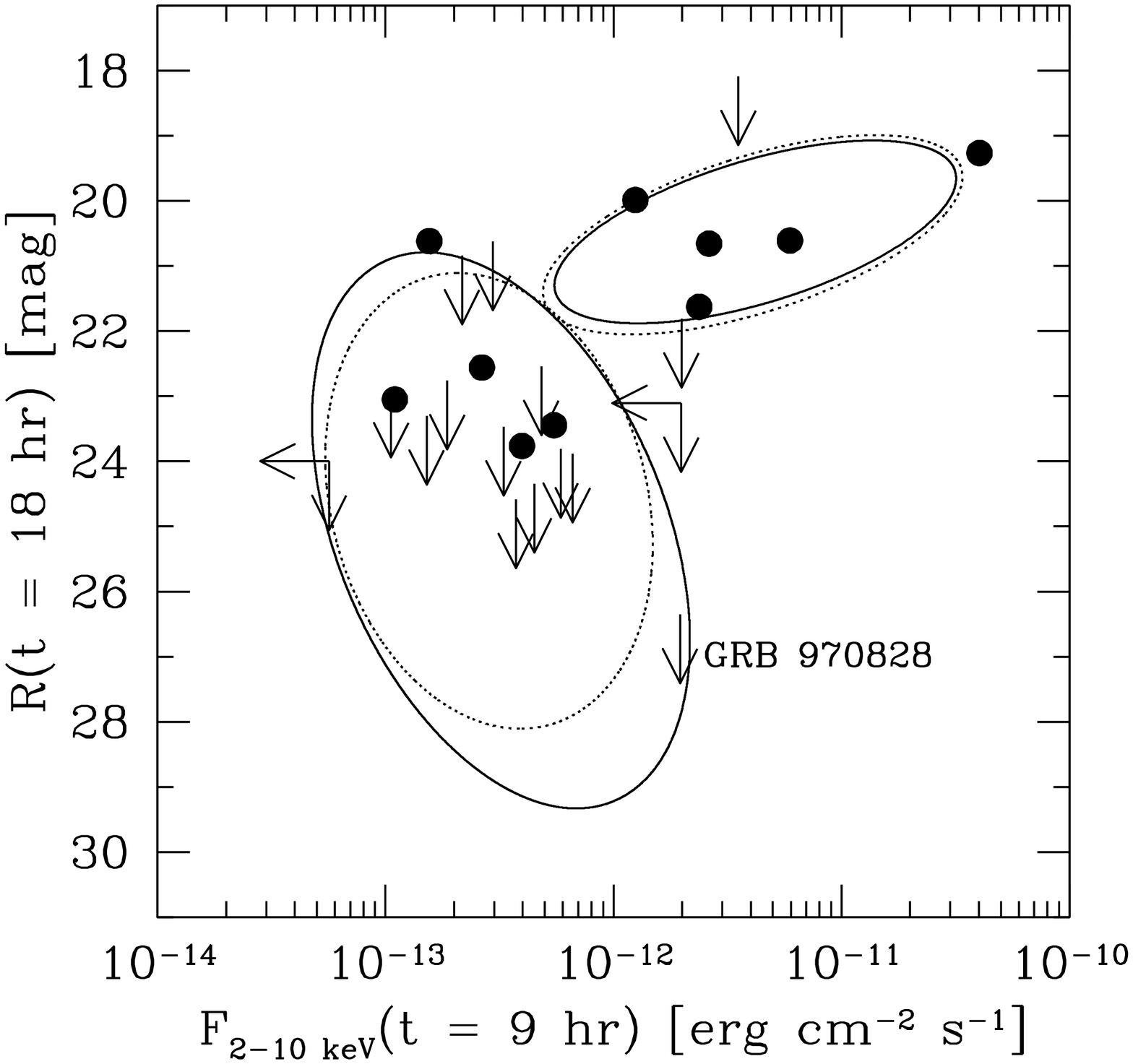]{R-band magnitude or limiting magnitude, scaled to 18 hours after the burst, vs. 2 -- 10 keV flux or limiting flux, scaled to 9 hours after the burst.  The solid ellipse is from the best-fit model of \S 5 to all of the data, and the dotted ellipse is from the best-fit model to all of the data excluding GRB 970828.  The ellipses should encompass $\approx 90$\% of the data.  The optical and X-ray data appear to be in good, qualitative agreement with the theoretical expectation of Figure 6, and clearly GRB 970828 does not dominate the fit.\label{rxdist2.eps}}

\figcaption[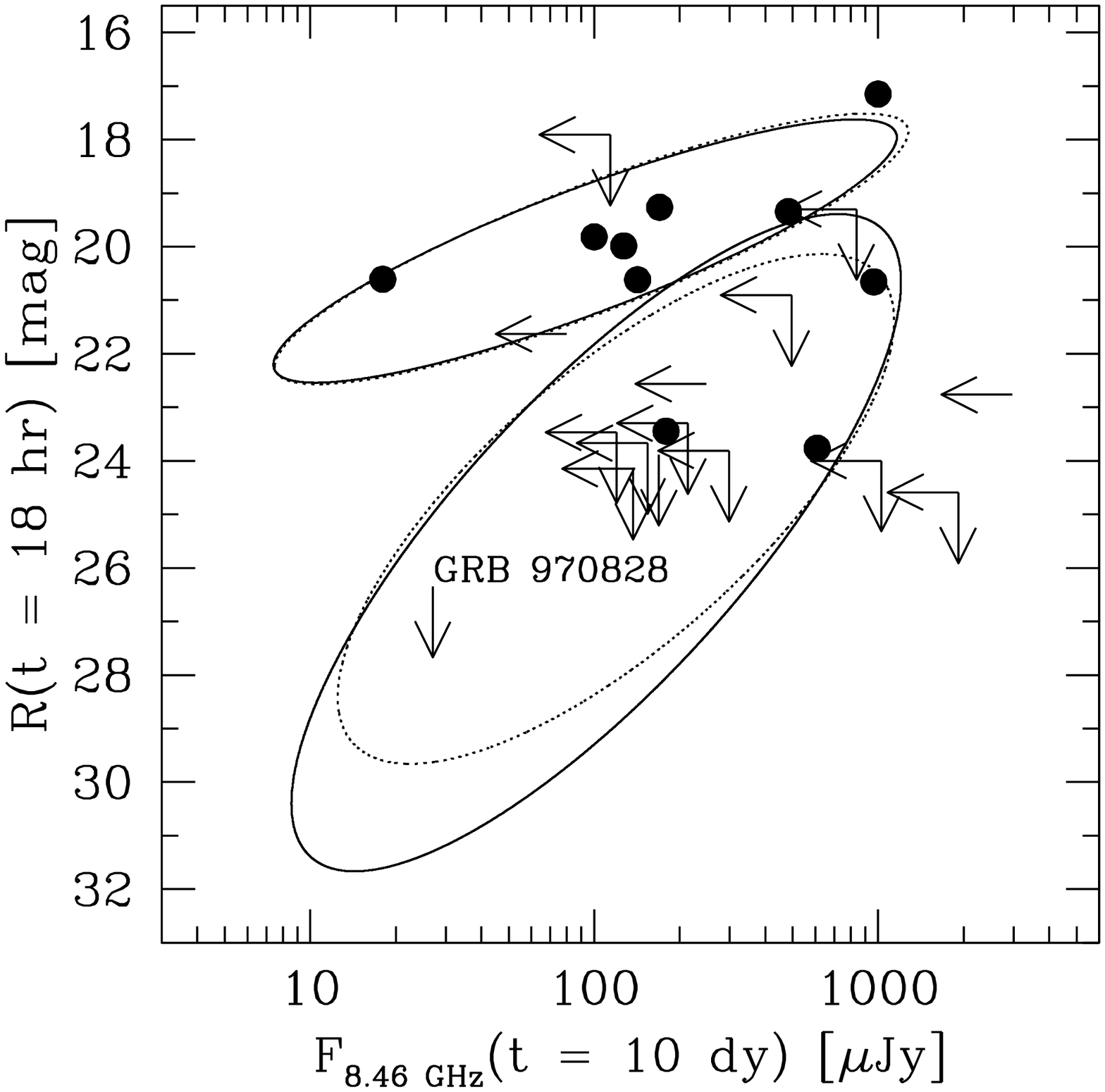]{R-band magnitude or limiting magnitude, scaled to 18 hours after the burst, vs. 8.46 GHz flux density or limiting flux density, scaled to 10 days after the burst.  The solid ellipse is from the best-fit model of \S 5 to all of the data, and the dotted ellipse is from the best-fit model to all of the data excluding GRB 970828.  The ellipses should encompass $\approx 90$\% of the data.  The optical and radio data appear to be in good, qualitative agreement with the theoretical expectation of Figure 7, and clearly GRB 970828 does not dominate the fit.\label{rrdist2.eps}}

\clearpage

\setcounter{figure}{0}

\begin{figure}[tb]
\plotone{rdist.eps}
\end{figure}

\begin{figure}[tb]
\plotone{rxdist.eps}
\end{figure}

\begin{figure}[tb]
\plotone{lc828.eps}
\end{figure}

\begin{figure}[tb]
\plotone{lc123.eps}
\end{figure}

\begin{figure}[tb]
\plotone{rrdist.eps}
\end{figure}

\begin{figure}[tb]
\plotone{rxdist3.eps}
\end{figure}

\begin{figure}[tb]
\plotone{rrdist3.eps}
\end{figure}

\begin{figure}[tb]
\plotone{rxdist2.eps}
\end{figure}

\begin{figure}[tb]
\plotone{rrdist2.eps}
\end{figure}

\end{document}